\documentclass[12pt]{iopart}
\newcommand{\sect}[1]{\setcounter{equation}{0}\section{#1}}

\newcommand{\bfm}[1]{\mbox{\boldmath${#1}$}}

\newcommand{\half}{\frac{1}{2}}
\newcommand{\adag}{a^{\dag}}
\usepackage{dcolumn}
\begin{document}

\title[$q$-Interacting particle systems]{Interacting particles system revisited in the
framework of the $\bfm q$-deformed algebra}
\author{A.M. Scarfone$^1$, and P. Narayana Swamy$^2$}
\address{$^1$Dipartimento di Fisica and Istituto Nazionale per la Fisica della Materia (CNR--INFM),
Sezione del Politecnico di Torino, I-10129, Italy}
\address{$^2$Department of Physics, Southern
Illinois University, Edwardsville, IL 62026, U.S.A.}
\date{\today}

\begin {abstract}
We discuss the possibility of interpreting a $q$-deformed
non-interacting system as incorporating the effects of
interactions among its particles. This can be accomplished, for
instance, in an ensemble of $q$-Bosons by means of the virial
expansion of a real gas in powers of the deformed parameter. The
lowest order virial coefficient reduces to the case of the
standard, non-interacting Bose gas, while the higher order virial
coefficients contain effects arising from the interaction. The
same picture can be drawn in a quantum mechanical system where it
is shown that the $q$-deformed momentum can be expanded in a
series contains high-order powers of the standard quantum
phase-space variables. Motivated by this result, we introduce,
in the classical framework, a transformation relating the momentum
of a free system with the momentum of an interacting system. It is
shown that the canonical quantization applied to the
interacting system imply a $q$-deformed quantization for the free system.
\end {abstract}
\submitto{JPA} 

\eads{\mailto{antonio.scarfone@polito.it,
pswamy@siue.edu}} \maketitle

\sect{Introduction} The earliest studies to develop a new
realization of the quantum group SU$_q(2)$ are perhaps those of
Biedenharn and Mcfarlane \cite{Biedenharn,g1}. This construction,
which is now a standard procedure in the literature, is based on
the following algebra describing a $q$-Bosons oscillators system
\begin{eqnarray}
\nonumber
&&\tilde a\,\tilde\adag-q\,\tilde\adag\,\tilde a=1 \ ,\quad [\tilde
a,\,\tilde a]=0 \ ,\quad[\tilde\adag,\,\tilde\adag]=0 \ ,\\
&&[N,\,\tilde a]=-\tilde a \ ,\quad [N,\,\tilde\adag]=\tilde\adag \
,\label{1}
\end{eqnarray}
where $q$ is a real deformation parameter. The $q$-Boson
Hamiltonian operator is
$H=\half\,\omega\,(\tilde\adag\,\tilde a+\tilde
a\,\tilde\adag)$
(hereinafter we use unities with $\hbar=1$),
where eigenvalues $E_n=\half\,\omega\,([n+1]+[n])$ are
expressed in terms of the basic numbers $[n]=(q^n-1)/(q-1)$ and the
Fock space of the $q$-Boson states is constructed according to
$|n\rangle=(1/\sqrt{[n]!})\,(\adag)^n\,|0 \rangle$,
where the $q$-factorial is given by $[n]!=[n]\,[n-1]\ldots[1]$, when
non-zero.

Starting from the above formalism, the statistical
mechanics of a $q$-Bosons gas has been investigated rather
extensively in literature \cite{g2,g3,g4,g5,g6}. Recent studies on
the nature of the thermodynamics as well as properties relating to
the algebra and Heisenberg's equation of motion have been worthwhile
\cite{PNS}. More recently, specific topics relating to the
thermostatistics of $q$-deformed Boltzmann, Bose and Fermi systems
have been investigated in detail \cite{ALAMSPNS-1,ALPNS-1}. These
studies, have not only established detailed results for the
thermostatistical properties of ideal $q$-Bosons and $q$-Fermions
systems, but they have also illustrated how the deformation of
classical systems can be described by means of the $q$-exponential
functions and the concomitant Fokker-Planck equation \cite{LSS}. In
addition, the study of $q$-Bosons and $q$-Fermions is particularly
instructive since it has been shown that the $q$-calculus based on
the Jackson derivative can be successfully used to modify standard
thermodynamic relations in  such a way that the theory of these ideal
$q$-deformed systems can be formulated in a self-consistent manner.

Moreover, classical systems described by the $q$-deformed Poisson
bracket has been investigated \cite{lss1} in order to
understand the origin of the deformed structure from the
underlying dynamics governing
some complex systems.

Having accomplished a great deal of understanding of the
$q$-deformed systems revealed by recent investigations, we may ask
the question whether there are simple physical descriptions of
what constitutes a deformation. In other words, it is worthwhile
to ask if there are simple manifestations of $q$-deformation and
what are the physical basis of such manifestations. It has
been observed in \cite{FW} that interacting Bosons gas can be
described by a Hamiltonian containing various powers of the
operators $a$ and $\adag$, where the lowest order non-interacting
Hamiltonian contains only one pair of these operators.
Furthermore, we can learn from a study of non-ideal gases, (see,
for instance, \cite{LL}), that the state equations of interacting
particle systems are well described by means of the virial
expansion in terms of higher orders by  powers of the number of
particles, where the Van der Walls
approximations is given by the first two terms in the series.

Therefore, the power series expansion is a convenient tool to
describe interacting systems, with the lowest order representing the
system of non-interacting particles in the ensemble. Unfortunately,
very few authors have examined this issue. An exception
is given by Parthasarathy and Viswanathan  \cite{RP-KV} who have
developed an interesting interpretation of the origin of the
$q$-deformation in an interacting particles system. In fact, it is
shown that many thermodynamic quantities of a $q$-Bosons system can
be expressed as a power series in powers of $N$ or in series of the
deformed parameter $\epsilon=q-1$, implying that the $q$-deformation
arises from the interaction among the particles of the ensemble.
More precisely, while the zero-th order describes the undeformed (non-interacting) system, the
higher order terms contain various powers of $\epsilon$ which are
equivalent to different order contributions arising from interactions.

Similar considerations have been presented also in \cite {Zhang},
where the perturbative aspects of the $q$-deformed Schr\"odinger
equation are analyzed by means of a high order momentum-dependent
interaction originating from the underlying $q$-deformation of the
Heisenberg algebra.

In this paper, we shall highlight some aspects concerning the
possible interpretation of a $q$-deformed non-interacting
system describing a non-deformed interacting system.

The plan of the paper is as follows. In Section II, we present
some results to illustrate the idea that $q$-deformation can
originate from the presence of an interaction in the system. This
is better clarified in Section III, where we study an interacting
Bosons system by means of the virial expansion of a free $q$-Bosons
system. In Section IV, we illustrate the same idea for a quantum
mechanical system, by revisiting the $q$-deformed Heisenberg
algebra. We write down the $q$-deformed momentum in a series of
the standard quantum space-phase variables where the high-order
contributions originate from the presence of the interaction into
the system. The equivalence of $q$-deformation and interaction for
a mechanical system is discussed in the Section V by introducing a
contact transformation which relates the classical counterpart of
the canonical quantum momentum (describing an interacting
system) to the classical counterpart of the deformed quantum
momentum (describing a free system). As a consequence, it is shown
that the canonical quantization ({\em \`{a} la Dirac}) applied to
the interacting system implies a $q$-deformed quantization of the
free system. Section VI contain a conclusive discussion.


\sect{Preliminary results}

We consider a system of interacting particles described by the
$q$-oscillator algebra, expressed for a one level system through equation
(\ref{1}), where $q$ implements the deformation of the system. The
theory of such $q$-Bosons introduces the basic number according to
\begin{equation}
[N]=\frac{q^N-1}{q-1} \ ,\label{2}
\end{equation}
where $N$ is the number operator,
in order to describe the Fock space of the particles.\\
Hereinafter, we take $q>1$ without loss of generality and we set
$\epsilon=q-1$, with $\epsilon \ll 1$. Thus, the expression of the
basic number is written as
\begin{equation}
[N]=\epsilon^{-1}\,\Big[(1+\epsilon)^N-1\Big] \ ,\label{3}
\end{equation}
which can be expanded in a series of powers of $N$ according to
\begin{eqnarray}
\nonumber
[N]&=&\left(1-{\epsilon\over2}+{\epsilon^2\over3}-{\epsilon^3\over4}
+\ldots\right)\,N
+{\epsilon\over2!}\,\left(1-\epsilon+{11\over12}\,\epsilon^2-\ldots\right)\,N^2\\
& &+{\epsilon^2\over3!}\,\left(1-{3\over2}\,\epsilon+\ldots\right)\,N^3+O(N^4)
\ .\label{4}
\end{eqnarray}

This result, in agreement with the interpretation advanced in
\cite{RP-KV}, shows that the quantity $[N]$ incorporates thermal
averages of $N,\,N^2,\,N^3$, etc. Further inspection reveals that
the contributions from interaction can be viewed either in terms of
$N, N^2, N^3$, etc. or in terms of $\epsilon,\,\epsilon^2,
\,\epsilon^3$, etc. In fact, by interpreting the contributions
arising from interactions by means of terms
containing higher powers of the deformation parameter $\epsilon$, we can
rewrite the series (\ref{4}) in the following form
\begin{eqnarray}
\nonumber
[N]&=&\sum_{i=0}^\infty{\epsilon^i\over(i+1)!}\,\prod_{j=0}^i(N-j)\\
&=&N+{\epsilon\over2!}\,N\,(N-1)+{\epsilon^2\over3!}\,N\,(N-1)\,(N-2)
+O(\epsilon^3) \ ,
\label{5}
\end{eqnarray}
so that, we may regard the above series as describing the
interaction among the particles of the system: the $\epsilon
\rightarrow 0$ limit describes the non-interacting particles
corresponding to an undeformed gas while the higher order terms
describe the various orders of interaction. In this way, we can
interpret the system of interacting particles  as a $q$-deformed
free gas.

Next, we consider the standard form of the Hamiltonian
for the $q$-Bosons
\begin{equation}
H_\epsilon={1\over2}\,\omega\,\Big([N+1]+[N]\Big) \ ,\label{7}
\end{equation}
reducing to the usual Hamiltonian
$H_0=\omega\,(N+1/2)$ when $\epsilon$ goes to zero. This Hamiltonian follows directly from the $q$-deformed oscillator
algebra, with the standard representation $[N]=\adag\,a$ and
$[N+1]=a\,\adag$.\\ The alternative form
$H_\epsilon=\omega\,\Big([N]+1/2\Big)$ is also commonly used
\cite{Chaichian}, but that is certainly an approximation \cite{noi}.

Hamiltonian (\ref{7}) can be expressed as
\begin{eqnarray}
\nonumber
H_\epsilon&=&\omega\,\left[{1\over2}+\left(1+{\epsilon^2\over12}-{\epsilon^3\over12}+\ldots\right)\,N+
{\epsilon\over2}\,\left(1-{\epsilon\over2}+{5\over12}\,\epsilon^2-\ldots\right)\,N^2\right.\\
& &\left.+{\epsilon^2\over6}\,
\left(1-\epsilon+\ldots\right)\,N^3\right]+O(N^4) \ ,\label{8}
\end{eqnarray}
which can be rearranged in a series containing higher powers of the
deformation parameter according to
\begin{equation}
H_\epsilon=H_0+\omega\sum_{i=1}^\infty\epsilon^i\,{(2\,N+1-i)\over2\,(i+1)!}\,
\prod_{j=0}^{i-1}(N-j) \ .
\end{equation}
The first term $H_0$, surviving in the limit $\epsilon \rightarrow 0$,
is the undeformed Hamiltonian corresponding to
the non-interacting case, whilst the higher order terms are interpreted
 as arising from interaction.

Finally, for completeness, let us consider the following transformation
\begin{equation}
\tilde a=\hat a\,\sqrt{[N]\over N} \ ,\label{9}
\end{equation}
and similarly for $\tilde\adag$, where $\hat a$ and $\hat\adag$
are the undeformed generators obeying to the ordinary SU(2) algebra. As is
well-known \cite{Kulish,Swamy2}, this transformation maps the
$q$-algebra (\ref{1}) onto the standard algebra of a non
interacting Bosons system. By
expanding equation (\ref{9}) in power series, thus one obtains
\begin{eqnarray}
\nonumber \tilde a=\hat
a\,\Big[1&+&{\epsilon\over4}\,(N-1)-{\epsilon^2\over96}\,(N-1)\,(5\,N+3)\\
&+&
{\epsilon^3\over384}\,(N-1)\,(N-5)\,(3\,N-7)+O(\epsilon^4) \Big] \
, \label{10}
\end{eqnarray}
which shows the equivalence of the two operators $\tilde a$ and
$\hat a$ at the zeroth order. Again, the high order terms can be
interpreted as a correction to $\hat a$ arising from the presence of
the interactions in the system of non-interacting Bosons.


\sect{The virial expansion in the $\bfm q$-Bosons system}

A good illustration of the idea that a $q$-deformed non-interacting
many-body system can be interpreted as describing an non-deformed
interacting system can be provided by studying the virial expansion
of a $q$-Bosons ensemble although this idea can also be investigated
starting from other physically relevant quantities. For instance,
one could consider the probability distribution function for a
$q$-Bosons system, derived in \cite{ALPNS-1}, as
\begin{equation}
n={1\over\ln(1-\epsilon)}\,\ln\left({y\over y-\epsilon}\right) \ ,\label{11}
\end{equation}
where $y=z^{-1}\,e^{\beta\,E}-1$ and $z=e^{\beta\,\mu}$ is the
fugacity of the gas. By expanding this quantity in a power series,
one obtains
\begin{eqnarray}
\nonumber
n={1\over y}&+&{\epsilon\over2\,y}\,\left(1+{1\over y}\right)
-{\epsilon^2\over12\,y}\,\left(1+{1\over y}\right)\,\left(1-{4\over
y}\right)\\
&+&{\epsilon^3\over24\,y}\,\left(1+{1\over
y}\right)\,\left(1-{2\over y}+{6\over y^2}\right)+O(\epsilon^4) \
,\label{12}
\end{eqnarray}
where the first term corresponds to the mean occupation number of
the standard Bose statistics, whilst the higher order terms
represent interaction contributions. We may thus interpret the
above as the average occupation number of an interacting ensemble
of Bosons.

Most results derived in a thermostatisics theory
based on the $q$-calculus involve generalized functions such as
$\tilde g_{3/2}(z,\,\epsilon)$ or $\tilde g_{5/2}(z,\,\epsilon)$.
These functions are generalizations of the well-known functions
$g_{3/2}(z)$ or $g_{5/2}(z)$ arising in the standard
thermostatistics of a Bosons gas and can be taken as typical of
what comes out in the deformed theory which we are now
interpreting as consequence of the interaction among the many-body
system.\\ We start from the following result derived in
\cite{ALPNS-1}
\begin{equation}
\tilde g_{3/2}(z,\,\epsilon)=\frac{\lambda^3}{v} \ ,\label{B1}
\end{equation}
where $v=V/N$ is the specific volume, $\lambda = h/(2\,\pi\,m\,k\,T)^{\half}$ is the thermal wavelength and
\begin{equation}
\tilde
g_{3/2}(z,\,\epsilon)=\frac{1}{\ln(1+\epsilon)}\sum_{k=1}^{\infty}
\frac{z^k}{k^{5/2}}\,\left[(1+\epsilon)^k-1\right] \ .\label{B1-1}
\end{equation}
Expanding this function in a power series, we obtain the result
\begin{eqnarray}
\nonumber & &\hspace{-17mm}\tilde
g_{3/2}(z,\,\epsilon)=z\,\left(1+\frac{z}{2\sqrt{2}}+\frac{z^2}{3\sqrt{3}}+\ldots
\right)+\frac{z}{2}\,\left(1+\frac{z}{\sqrt{2}}+\frac{z^2}{\sqrt{3}}
+\ldots\right)\,\epsilon\\
\nonumber
& &-\frac{z}{12}\,\left(1-\frac{z}{\sqrt{2}}-\sqrt{3}\,z^2
+\ldots\right)\,\epsilon^2+\frac{z}{24}\,\left(1+\frac{z^2}{\sqrt{3}}
+\ldots\right)\,\epsilon^3 +O(\epsilon^4) \ .\\ \label{13}
\end{eqnarray}
Note that the zeroth order term is the familiar $g_{3/2}(z)$ function encountered in standard Bose statistics \cite{Pathria} whilst the higher order terms can be interpreted as describing
contributions from interactions.

In a similar way, we can consider the equation of state for a
$q$-Bosons system \cite{ALPNS-1}
\begin{equation}
\tilde g_{5/2}(z,\,\epsilon)=\frac{\lambda^3\,P}{k\,T} \ ,\label{B10}
\end{equation}
where the function
\begin{equation}
\tilde
g_{5/2}(z,\,\epsilon)=\frac{1}{\ln(1+\epsilon)}\sum_{k=1}^{\infty}\frac{z^k}{k^{7/2}}\,\left[
(1+\epsilon)^k-1\right] \ ,\label{B11}
\end{equation}
has the following power series expansion
\begin{eqnarray}
\nonumber & &\hspace{-17mm}\tilde
g_{5/2}(z,\,\epsilon)=z\,\left(1+\frac{z}{4\,\sqrt{2}}+\frac{z^2}{9\,\sqrt{3}}
+\ldots\right)+\frac{z}{2}\,\left(1+\frac{z}{2\,\sqrt{2}}+
\frac{z^2}{3\,\sqrt{3}}+\ldots\right)\,\epsilon\\
\nonumber&
&-\frac{z}{12}\,\left(1-\frac{z^2}{2\,\sqrt{2}}-\frac{z^3}{\sqrt{3}}
+\ldots\right)\,\epsilon^2+\frac{z}{24}\,\left(1+\frac{z^2}{3\,\sqrt{3}}
+\ldots\right)\,\epsilon^3+O(\epsilon^4) \ .\\ \label{B12}
\end{eqnarray}
Again, the zeroth order term coincides with the known function $g_{5/2}(z)$
of the standard statistical mechanics, whilst the higher order
terms can be interpreted as describing contributions from
interactions.

From the results (\ref{13}) and (\ref{B12}) we can
obtain the following virial expansion
\begin{equation}
\frac{P\,v}{k\,T}=\sum_{k=1}^{\infty}a_k(\epsilon)\,\left(\frac{\lambda^3}{v}
\right)^{k-1} \ ,\label{B13}
\end{equation}
where the virial coefficients $a_k(\epsilon)$ describe the equation
of state of an interacting Bosons system or, equivalently, they
describe the equation of state of a $q$-deformed system of
non-interacting Bosons. The first few terms are
\begin{eqnarray}
&&\hspace{-17mm}a_1(\epsilon)=1 \ ,\label{B14}\\
\nonumber&&\hspace{-17mm} a_2(\epsilon)=-{1\over4\,\sqrt{2}}-{1\over48\,
\sqrt{2}}\,\epsilon^2\,(1-\epsilon)+O(\epsilon^4)\\
&&\hspace{-8mm}=-0.17678-0.01473\,\epsilon^2+0.01473\,\epsilon^3+O(\epsilon^4) \ ,\label{B15}\\
\nonumber
&&\hspace{-17mm}a_3(\epsilon)=-\left({2\over9\,\sqrt{3}}-{1\over8}\right)
-\left({1\over18\,\sqrt{3}}-{1\over48}\right)\,\epsilon^2\,(1-\epsilon)+O(\epsilon^4)
\\
&&\hspace{-8mm}=-0.00330-0.01124\,\epsilon^2+0.01124\,\epsilon^3+O(\epsilon^4) \
,\label{B16}\\
\nonumber
&&\hspace{-17mm}a_4(\epsilon)=-\left({3\over32}+{5\over32\,\sqrt{2}}
-{1\over2\,\sqrt{6}}\right)-\left({3\over64}+{5\over128\,\sqrt{2}}-
{1\over6\,\sqrt{6}}\right)\,\epsilon^2\,(1-\epsilon)+O(\epsilon^4)
\\
&&\hspace{-8mm}=-0.00011-0.00645\,\epsilon^2+0.00645\,\epsilon^3+O(\epsilon^4) \
.\label{B17}
\end{eqnarray}
(We may note that the first coefficient is exact since it does not
contain any corrections).

It is readily observed that when
$\epsilon\to0$ we recover the virial coefficients of the ordinary
(undeformed) non-interacting Bosons system, as shown in standard
textbooks \cite{Pathria}. The higher order terms arise from the
deformation, i.e. we may regard the deviations from the values of
the ordinary Bosons system as indicating the presence of suck an
interaction among the particles of the system.


\sect{Heisenberg algebra and $\bfm q$-deformed momentum}

We may further develop the idea of the equivalence among
deformation and interaction by considering a quantum mechanical system.

As  is known, a (non-Hermitian realization) of algebra
(\ref{1}) is provided by means of the $q$-deformed Bargmann-Wigner
holomorphism, given by $\tilde a\equiv x$ and $\tilde\adag\equiv
D_x$, where $D_x$ is the Jackson derivative \cite{Jackson},
defined in
\begin{equation}
D_x={1\over x}\,\frac{{\cal D}_q-1}{q-1} \ ,\label{A1}
\end{equation}
with
\begin{equation}\label{A2}
{\cal D}_q=q^{x\,\partial_x} \ ,
\end{equation}
the dilatation operator
\begin{equation}
{\cal D}_q\,f(x)=f(q\,x)\,{\cal D}_q \ .
\end{equation}
According to the $q$-algebra (\ref{1}), by taking into account the relation
$[N+1]=q\,[N]+1$, we can derive the following solution \cite{Finkelstein1,Finkelstein2}
\begin{equation}
[N]=x\,D_{x} \ ,\label{N}
\end{equation}
so that the $q$-Boson number operator can be defined as $N=x\,\partial_{x}$ [see equation (\ref{A1})].

With the purpose of  consistently introducing a quantum mechanics
theory based on the $q$-calculus, it has been suggested by several
authors (see for instance \cite{Dobrogowska,Dayi,Minahan}),
to replace the quantum momentum operator $\bar p=-i\,\partial_x$
with a $q$-deformed version given by
$\tilde P=-i\,D_{\tilde X}$.\\ In this way, the algebra (\ref{1}) can be
written formally in the form
\begin{equation}
q\,\tilde X\,\tilde P-\tilde P\,\tilde X=i \ ,\label{a3}
\end{equation}
with $\tilde X\equiv x$ the position operators, which reduces to
the standard Heisenberg algebra in the $q\to1$ limit.\\ Despite this
nice result a certain prudence is indeed required. In fact, while
$\bar p$ is a Hermitian operator and can be
identified with a physical observable, the same is not true for
the quantity $\tilde P$ which is not Hermitian. Thus, a physical
observable corresponding to a $q$-deformed momentum
$\tilde p$ \cite{Fichtmuller,Cerchiai} must be introduced
according to a certain procedure such that $\tilde p=\tilde
p^\dag=f(\tilde P,\,\tilde P^\dag)$.

To do this, we first
observe that the ordinary commutation among the operators $\tilde
X$ and $\tilde P$ reads
\begin{equation}
\tilde X\,\tilde P-\tilde P\,\tilde X=i\,{\cal D}_q \ .\label{a5}
\end{equation}
Of course, equations (\ref{a3}) and (\ref{a5}) represent the same
algebra, as can be verified by employing the relation ${\cal
D}_q=1+i\,(q-1)\,\tilde X\,\tilde
P$.\\
On the other hand, let us consider the Hermitian conjugate of this
last equation, given by
\begin{equation}
\tilde X\,\tilde P^\dag-\tilde P^\dag\,\tilde X=i\,\left(q\,{\cal
D}_q\right)^{-1} \ ,\label{h1}
\end{equation}
since ${\cal D}_q^\dag=(q\,{\cal D}_q)^{-1}$ and we assumed $\tilde
X=\tilde X^\dag$. By matching equation (\ref{a3}), multiplied on the left hand
side by the operator $(q\,{\cal D}_q)^{-1}$ that can be written in
\begin{equation}
\tilde X\,\left(q\,{\cal D}_q\right)^{-1}\,\tilde P-\left(q\,{\cal
D}_q\right)^{-1}\, \tilde P\,\tilde X=i\,\left(q\,{\cal
D}_q\right)^{-1} \ ,\label{h2}
\end{equation}
with equation (\ref{h1}), it is natural to define
\begin{equation}
\tilde P^\dag=\left(q\,{\cal D}_q\right)^{-1}\,\tilde P \ , \
\quad{\rm with}\quad \left(\tilde P^\dag\right)^\dag\equiv \tilde P
\ .\label{hermit}
\end{equation}
Accordingly, one introduces the Hermitian operator
\begin{equation}
\tilde p={1\over2}\left(\tilde P+\tilde P^\dag\right) \ ,\label{hp}
\end{equation}
which can be identified with the right $q$-deformed version of the quantum
 momentum.\\ After rescaling the position operator,
according to $\tilde x=2\,q\,\tilde X/(1+q)$, algebra (\ref{a3}) can
be finally written in
\begin{equation}
q^{1/2}\,\tilde x\,\tilde p-q^{-1/2}\,\tilde p\,\tilde
x=i\,\Lambda_q \ ,\label{hais1}
\end{equation}
where the operator $\Lambda_q=q^{-1/2}\,{\cal D}_q^{-1}$ fulfills
the further relations
\begin{equation}
\Lambda_q\,\tilde x=q^{-1}\,\tilde x\,\Lambda_q \ ,\quad\Lambda_q
\,\tilde p=q\,\tilde p\,\Lambda_q \ .\label{hais2}
\end{equation}
Equations (\ref{hais1})-(\ref{hais2}) establish the $q$-deformed
Heisenberg's algebra for the phase space quantum variables
$\tilde x$ and $\tilde p$, as known in literature
\cite{Fichtmuller,Cerchiai}.

We can now consider a power series expansion for $\tilde p$. To do
this, we collect together Eqs. (\ref{A1}), (\ref{hermit}) and
(\ref{hp}) to obtain
\begin{equation}
\tilde p=-{i\over2\,\tilde X}\,{{\cal D}_q-{\cal D}_q^{-1}\over q-1}\equiv-{i\over\hat x}\,{q^{i\,\hat x\,\hat p}-q^{-i\,\hat x\,\hat p}\over q-q^{-1}} \ ,\label{pt}
\end{equation}
with $\hat x\equiv\tilde x$, $\hat p=-i\,\partial_{\hat x}$, which can be written in the following compact form
\begin{equation}
\tilde p=\hat p\,{[\hat z]_{\rm s}\over\hat z} \ ,
\end{equation}
where $\hat z=i\,\hat x\,\hat p$ and $[\hat z]_{\rm s}=(q^{\hat z}-q^{-\hat z})/(q-q^{-1})$ is the symmetric version of the basic number. Relation (\ref{pt}) is equivalent to the one derived in \cite{Fichtmuller} and successively reconsidered in \cite{Zhang}. It plays the same role
as the  transformation (\ref{9}) and relates the $q$-deformed
operators $\tilde x$ and $\tilde p$, fulfilling the generalized
Heisenberg's algebra (\ref{hais1}), to the standard operators $\hat x$
and $\hat p$ which satisfy the
undeformed algebra $[\hat x,\,\hat p]=i$.

By expanding equation (\ref{pt}) in a series of $\epsilon=q-1$,
we obtain
\begin{equation}
\tilde p=\hat p\,\left[1-{1\over6}\,\left(1+\hat x\,\hat
p\,\hat x\,\hat p\,\right)\,\epsilon^2\,(1-\epsilon)\right]+O(\epsilon^4) \ ,
\end{equation}
in accordance with the results reported in \cite{Zhang}.
Again, the higher order terms can be interpreted as effects due to
interactions, i.e. the first term corresponds to the momentum of an
non-interacting system whilst the higher order terms arise from
interactions.


\sect{The equivalent interaction}

In order to better clarify the question concerning the possible
link between interaction and deformation we are asking if there
exists a simple physical indication of what constitutes a deformation.

In this respect, we consider a classical system described by the
phase-space variables $(X,\,P)$. For a non-interacting system the
energy is only  kinetic, i.e. $E=P^2/(2\,m)$. Afterwards, we
introduce a contact transformation which changes the momentum
$P\to p\equiv p(X,\,P)$ but leave unchanged the coordinate $X\to
x\equiv X$. In the new phase-space variables $(x,\,p)$, the energy
of the system can be written, without loss of generality, in
$E=p^2/(2\,m)+U_\alpha(x,\,p)$ that contains now the extra term
$U_\alpha(x,\,p)$ representing an interaction potential, whose
explicit form is determined by the contact transformation we are
introducing. In particular, we are looking for a transformation
able to capture the physical meaning of the $q$-deformation.\\ On
the basis of the results described in the previous section, this
can be realized by introducing a transformation such that the
quantum analogue of the classical mapping $P\to p\equiv p(X,\,P)$
corresponds to the transformation (\ref{pt}) among quantum
operators. Hereafter, we  identify the quantity $\hat p$ with the
quantum operator of the momentum $p$ (describing the interacting
system) and the quantity $\tilde p$ with the quantum operator
corresponding to the momentum $P$ (describing the free system).

To realize such a result, we introduce the following point
transformation
\begin{equation}
x\to X\equiv x \ ,\quad p\to P={1\over x}\,{\sin(\alpha\,x\,p)\over
\sinh(\alpha)} \ ,\label{map}
\end{equation}
which imply an interaction term of the kind
\begin{equation}
U_\alpha(x,\,p)={\sin^2(\alpha\,x\,p)-\Big[\sinh(\alpha)\,x\,p\Big]^2\over2\,m\,\sinh^2(\alpha)\,x^2}
 \ ,\label{pot}
\end{equation}
[derived from the quantity $U_\alpha(x,\,p)={1\over2\,m}(P^2-p^2)$],
where $\alpha=\log q$ is the deformed parameter for the
classical system. In the $\alpha\to0$ limit equation (\ref{map})
reduces to the identity transformation and the potential $U_\alpha(x,\,p)$ vanishes.

It is easy to verify that, by employing the canonical quantum
prescription for the momentum $p$, with $p\to\hat
p=-i\,\partial_x$, we obtain a $q$-deformed quantum prescription
for the momentum $P$, with $P\to\tilde p$.\\ In fact, from
transformation (\ref{map}), we obtain
\begin{equation}
P={\sin(\alpha\,x\,p)\over
x\,\sinh(\alpha)}\quad\to\quad{1\over x}{\sin(-i\,\alpha\,x\,\partial_x)\over
\sinh(\alpha)}=-{i\over\hat x}\,{q^{i\,\hat x\,\hat p}-q^{-i\,\hat x\,\hat p}\over
q-q^{-1}}=\tilde p \ ,
\end{equation}
in accordance with the transformation (\ref{pt}).\\We remark that,
since transformation (\ref{map}) depends on both the phase-space
variables an ordering prescription must be imposed. Consistently
with the formalism adopted in the previous section, we employed
the standard prescription with the $\hat x$ operator on the left
and the $\hat p$ operator on the right.

An immediate consequence of this argument is that the
Schr\"odinger equation describing a quantum interacting system
obtained from the canonical quantization of the classical system
governed by the potential $U_\alpha(x,\,p)$ [$U_q(\hat x,\,\hat
p)={1\over2\,m}(\tilde p^2-\hat p^2)$ in the quantum picture] is transformed in a $q$-deformed
Schr\"odinger equation describing a non-interacting quantum system
(and vice versa).\\ This fact follows readily from the following
correspondence
\begin{equation}
E={\hat
p^2\over2\,m}+U_q(\hat x,\,\hat p)\quad\Leftrightarrow\quad E={\tilde p^2\over2\,m}  \ .
\end{equation}

Let us point out that the idea to introduce a deformed algebra in
order to take into account the effect of interactions is not new
at all.\\ In fact, this is reminiscent of what was done in early
quantum mechanics by Bohm and Madelung \cite{Bohm}, which
recognize that a quantum system, described by the non-commutative
operators $\hat x$ and $\hat p$ acting on the $\psi$-function of a
quantum system, behaves like a classical interacting fluid,
described by the commutative functions $\rho$ and $S$, whose
interaction is described through a suitable quantum
potential $U_{_{\rm BM}}(\rho)$, capable of capturing the quantum feature arising in the Schr\"odinger picture.\\
By identifying the quantum momentum $-i\,\partial_x\psi$ in the
$\psi$-representation with the canonical momentum $\partial_xS$ in the $\rho-S$
representation, the Schr\"odinger equation can be transformed according to
\begin{equation}
E=-{1\over2\,m}\,{\partial_x^2\psi\over\psi} \quad \Rightarrow
\quad E={(\partial_x\,S)^2\over2\,m}+U_{_{\rm BM}}(\rho) \ ,
\end{equation}
(endowed by a continuity equation for the field $\rho$), where, the transformed equation can
be interpreted as a classical Hamilton-Jacobi-like equation for the principal Hamilton function $S$.\\
In this way, the free system in the Schr\"odinger picture (in the noncommutative algebraic formalism) is
transformed in an interacting classical system (in the commutative algebraic formalism) governed by the Bohm-Madelung potential
\begin{equation}
U_{_{\rm BM}}(\rho)=-{1\over2\,m}{\partial_{xx}\,\sqrt\rho\over\sqrt\rho} \ ,
\end{equation}
that introduces the quantum effects
into the system just as  the potential (\ref{pot}) does with
respect to the $q$-deformation.

\sect{Conclusion and discussion}

In this work we have investigated a possible interpretation of
$q$-deformation in terms of a physical interaction. We analyze
several quantities of physical interest such as the Hamiltonian
and the occupation number of a $q$-Bosons system showing that all
these quantities can be expanded in a powers series of the
deformation parameter $\epsilon=q-1$. This leads us to identify
the zeroth order quantities corresponding to the description of an
undeformed system and the higher order quantities representing the
contribution arising from deformation. We can interpret the lowest
order quantities characterizing the non-interacting system whilst
the higher order contributions arise from the presence of
a kind of interaction in the system.

We investigate the virial expansion in the context of the
$q$-Bosons system which provides an illustration of the
equivalence of the two interpretations. This equivalence arises
from the premise that statistical mechanics reveals that a theory
of interacting particles, e.g., Bosons, predicts the various
virial coefficients so that the equation of state of the gas
reflects the presence of interactions in the system. Thus, we have
determined explicit forms for the first four virial coefficients
which are expressed in a powers series of $\epsilon$ and we have
shown explicitly that in the lowest order, these coefficients
reduce to those of the standard Bose gas, while  the higher order
terms describe the effects of the deformation. This idea stems
from a study of a non-ideal gas, (e.g. \cite{LL}), where the
equation of state of interacting systems is described well by the
virial expansion in powers of $N$, with the Van der Walls law
being just an approximate description provided
by the first two terms in the series.

The equivalence between $q$-deformation and interaction has been
also illustrated in a quantum mechanical system. The study of the
$q$-Heisenberg's algebra enables us to introduce the effective
momentum $\tilde p$ which can be related by a power series in the
ordinary quantum phase-space variables $\hat x$ and $\hat p$.
Based on this result we have considered a point transformation
$P\to p$ for the momentum of a classical system derived by
requiring that the canonical quantization rule $p\to \hat
p$, for themomentum $p$, imply a $q$-quantization rule $P\to\tilde
p$, for the momentum $P$. Accordingly,
identifying the $(X,\,P)$-system with a free system and the $(x,\,p)$-system
with the interacting one we have obtained the expression of the
interacting potential
$U_\alpha(x,\,p)$ responsible for the $q$-deformation.
This means that the classical momentum corresponding to the canonical quantum momentum (describing a suitably interacting
system) can be related to the classical momentum corresponding to the deformed
quantum momentum (describing a free system) by means of a contact
transformation.

At this stage we are not able to give a physical interpretation,
if any, of the potential (\ref{pot}). Notwithstanding, it is worth
remarking  that the idea underling this work, to interpret the
deformations of a physics theory as representative of the
interactions of the system that the theory is describing, has been also advanced by
several authors both for a quantum mechanical system \cite{Zhang}
and for a statistical mechanical system
\cite{ALAMSPNS-1,Quarati,Wang,Biro,Biro1} and actually is
applicable not just to the $q$-formalism employed in this work as a simple illustrative example.\\
Here, we have used
the $q$-deformed algebra, which is well known and largely applied
in physics since the beginning of the last century, in order to
give an explicit demonstration of how the method should work.
There is no physical reason, \emph{a priori}, to believe that the
$q$-deformation is the suitable deformation capable of capturing the
physical information contained in any interaction. More appropriately,
one should consider a given interacting system and then attempt to
realize the corresponding algebraic deformation able to capture the
physical feature contained in it. This project however, appears to be beyond our present scope.



\section*{References}

\begin{thebibliography}{99}

\bibitem{Biedenharn} Biedenharn L 1989 J. Phys. A: Math. Gen. {\bf22} L873

\bibitem{g1} Mcfarlane A 1989 J. Phys. A: Math. Gen. {\bf22} 4581

\bibitem{g2} Lee C and Yu J 1990 Phys. Lett. A {\bf150} 63

\bibitem{g3} Yang Q and Xu B 1993 J. Phys. A: Math. Gen. {\bf26} L365

\bibitem{g4} Aizawa N 1993 J. Phys. A: Math. Gen. {\bf26} 1115

\bibitem{g5} Viswanathan K {\em et al} 1993 J. Phys. A: Math. Gen. {\bf25} L335

\bibitem{g6} Ramanathan R 1992 Phys. Rev. D {\bf45} 4706

\bibitem{PNS} Narayana Swamy P 1996 Int. J. Mod. Phys. B
{\bf10} 683; 1996 Mod. Phys. Lett. B {\bf10} 23; 1998 Int. J.
Mod. Phys. B {\bf12} 3495; 2006 Int. J. Mod. Phys. B {\bf20} 697

\bibitem{ALAMSPNS-1} Lavagno A, Scarfone AM and Narayana
Swamy P 2007 J. Phys. A: Math. Theor. {\bf40} 8635

\bibitem{ALPNS-1} Lavagno A and Narayana Swamy P 2000
Phys. Rev. E {\bf61} 1218; 2002 Phys. Rev. E {\bf65} 036101

\bibitem{LSS} Lavagno A, Scarfone AM and Narayana Swamy P 2006
Eur. Phys. J. B {\bf50} 351

\bibitem{lss1} Lavagno A, Scarfone AM and Narayana Swamy P 2005
Rep. Math. Phys. {\bf55} 423; 2006 Eur. Phys. J. C {\bf47} 253

\bibitem{FW} Fetter A and Walecka J 1971
{\em Quantum Theory of many particle systems} (McGraw-Hill Book Company)

\bibitem{LL} Landau L and Lifshitz E 1980 {\em Statistical Physics} Vol 5 (Pergamon Press, New York)

\bibitem{RP-KV} Parthasarathy R and Viswanathan K 1992 preprint
IMSc-92/02-57

\bibitem{Zhang} Zhang J-z and Osland P 2001 Eur. Phys. J. C {\bf20} 393

\bibitem{Chaichian} Chaichian M, Gonzalez Felipe R and Montonen C 1993
J. Phys. A: Math. Gen. {\bf26} 4017

\bibitem{noi} Scarfone AM and Narayana Swamy P 2007 Int. J. Mod. Phys.
A {\bf22} 6169

\bibitem{Kulish} Kulish P and Damaskinsky E 1990 J. Phys. A: Math.
Gen. {\bf23} L415

\bibitem{Swamy2} Narayana Swamy P 2001 Mod. Phys. Lett. B {\bf15} 915

\bibitem{Pathria} Pathria RK 1972 {\em Statistical Mechanics}
(Pergamon Press, Oxford)

\bibitem{Jackson} Jackson FH 1909 Am. J. Math. {\bf38} 26

\bibitem{Finkelstein1} Finkelstein R 1998 Int. J. Mod. Phys. A {\bf13} 1795

\bibitem{Finkelstein2} Finkelstein R and Marcus E 1995 J. Math. Phys. {\bf36}
2652

\bibitem{Dobrogowska} Dobrogowska A and Odzijewicz A 2007 J. Phys. A: Math. Theor.
{\bf40} 2023

\bibitem{Dayi} Dayi \"{O}F and Duru IH 1997 Int. J. Mod. Phys. A {\bf12} 2373

\bibitem{Minahan} Minahan JA 1990 Mod. Phys. Lett. A {\bf5} 2625

\bibitem{Fichtmuller} Fichtm\"uller M, Lorek A and Wess J 1996 Z. Phys.
C {\bf71} 533

\bibitem{Cerchiai} Cerchiai BL, Hinterding R, Madore J and Wess J 1999
Eur. Phys. J. C {\bf8} 547

\bibitem{Bohm} Bohm D 1952 Phys. Rev. {\bf85} 166; {\em ibis} 1952
180

\bibitem{Madelung} Madelung E 1962 Z. Phys. {\bf40} 332

\bibitem{Quarati} Kaniadkis G, Quarati P and Scarfone AM 2002
Physica  A {\bf305} 76

\bibitem{Wang} Wang QA 2002 Eur. Phys. J. B {\bf26} 036142

\bibitem{Biro} B\'{\i}r\'{o} TS and Kaniadakis G 2006 Eur. J. Phys. B {\bf50}
3

\bibitem{Biro1} B\'{\i}r\'{o} TS and Purcsel G 2008 Phys. Lett. A {\bf372} 1174

\end{thebibliography}
\end{document}